\documentclass[aps,preprint,superscriptaddress,groupedaddress]{revtex4}  

\usepackage{graphicx}  
\usepackage{dcolumn}   
\usepackage{bm}        
\usepackage{amssymb}   
\usepackage{amsmath,textcomp}
\usepackage{xfrac}
\usepackage{natbib}
\usepackage[utf8]{inputenc}
\usepackage{float}

\begin{document}

\title{Observation of an Optical Spring With a Beamsplitter}

%

\author{Jonathan Cripe}
\affiliation{Department of Physics \& Astronomy, Louisiana State University, Baton Rouge, LA, 70808}

\author{Baylee Danz}
\affiliation{Department of Physics, Brigham Young University–Idaho, Rexburg, ID 83460}

\author{Benjamin Lane}
\affiliation{Department of Physics \& Astronomy, Louisiana State University, Baton Rouge, LA, 70808}

\author{Mary Catherine Lorio}
\affiliation{St. Joseph's Academy, Baton Rouge, LA, 70808}

\author{Julia Falcone}
\affiliation{Department of \& Astronomy, Case Western Reserve University, Cleveland, OH 44106}

\author{Garrett D. Cole}
\affiliation{Vienna Center for Quantum Science and Technology (VCQ), Faculty of Physics, University of Vienna, A-1090 Vienna, Austria}
\affiliation{Crystalline Mirror Solutions LLC and GmbH, Santa Barbara, CA, and Vienna, Austria}

\author{Thomas Corbitt}
\email{tcorbitt@phys.lsu.edu}
\affiliation{Department of Physics \& Astronomy, Louisiana State University, Baton Rouge, LA, 70808}




\begin{abstract}
We present the experimental observation of an optical spring without the use of an optical cavity. The optical spring is produced by interference at a beamsplitter and, in principle, does not have the damping force associated with optical springs created in detuned cavities. The experiment consists of a Michelson-Sagnac interferometer (with no recycling cavities) with a partially reflective GaAs microresonator as the beamsplitter that produces the optical spring. Our experimental measurements at input powers of up to 360 mW show the shift of the optical spring frequency as a function of power and are in excellent agreement with theoretical predictions. In addition, we show that the optical spring is able to keep the interferometer stable and locked without the use of external feedback. 
\end{abstract}




\maketitle

\section{Introduction}

Optomechanical cavities consisting of a moveable mirror or resonator allow the electromagnetic radiation of the cavity mode to couple to the motion of the mechanical oscillator. Optomechanical cavities have been proposed for improving the sensitivity of gravitational wave detectors below the standard quantum limit (SQL), tests of quantum mechanics, and quantum information \cite{24}.

One feature coupling light to a mechanical resonator in a cavity is the optical spring effect, which was first discussed for Fabry-P\'erot cavities by Braginsky \cite{Braginsky_1, Braginsky_2}. 
For the traditional case of the Fabry-P\'erot cavity, the optical spring is created in a detuned cavity where the cavity's circulating power, and therefore the radiation pressure force on the mirrors, is proportional to the cavity length \cite{13}. For a blue-detuned cavity in which the cavity's resonance frequency is less than the laser frequency, the linear relationship between the radiation pressure force and cavity length creates a positive restoring force with an effective spring constant $K_\mathrm{OS}$ and an antidamping force $\Gamma_\mathrm{OS}$. The combination of the optical spring constant and the mechanical spring constant of the device combine to shift the resonance  frequency of the system from $\Omega_\mathrm{m}$ to $\sqrt{\Omega_\mathrm{m}^2+\Omega_\mathrm{OS}^2}$ where $\Omega_\mathrm{m}$ is the resonance frequency of the mechanical oscillator and $\Omega_\mathrm{OS}$ is the optical spring frequency \cite{9, 13, Cripe_RPL}. This frequency shift is an experimental signature of the optical spring. 


The antidamping force created by the optical spring can overwhelm the mechanical damping and lead to dynamic instabilities \cite{2-2,15,16} and is usually controlled with feedback loops \cite{2-2,15, Cripe_RPL}. An alternative method to stabilizing the optical spring is to modify the damping force by adding a second optical spring \cite{17, Singh_PRL} or utilizing thermo-optic effects \cite{18, PA_thermal}.

Although the detuned  Fabry-P\'erot cavity is the canonical example of creating an optical spring, it is possible to create an optical spring in other topologies. An optical spring can be created in any system that is able to produce a linear relationship between the radiation pressure force and displacement. Dual-recycled gravitational wave detectors such as Advanced LIGO \cite{LIGO} and Advanced VIRGO \cite{VIRGO} are able to create an optical spring in the signal recycling cavity by detuning the signal recycling mirror \cite{Chen, 1-1}. A Michelson-Sagnac interferometer with a signal recycling mirror at the dark port can also produce an optical spring by changing the position of the signal recycling mirror \cite{25, 19}. 

These examples, however, still rely on the use of a cavity to produce the optical spring. In this paper, we present the measurement of an optical spring produced by the interaction of two input fields at a beamsplitter, which we will refer to as the microresonator, similar to the scheme outlined in \cite{Danilishin2012}. To achieve this, we utilize a Michelson-Sagnac interferometer for simplicity. Previous results using a Michelson-Sagnac interferometer have included a signal recycling mirror and have not directly observed the frequency shift that accompanies the optical spring \cite{19}.  We measure the optical spring at input powers of 50 mW, 100 mW, 200 mW, and 360 mW and compare our experimental results with a theoretical model. 
The optical springs created at all four input powers are strong enough to keep the interferometer stable and locked without the use of any external electronic feedback or additional optical fields.

\begin{figure}[]
\center
\includegraphics[width=0.5\columnwidth]{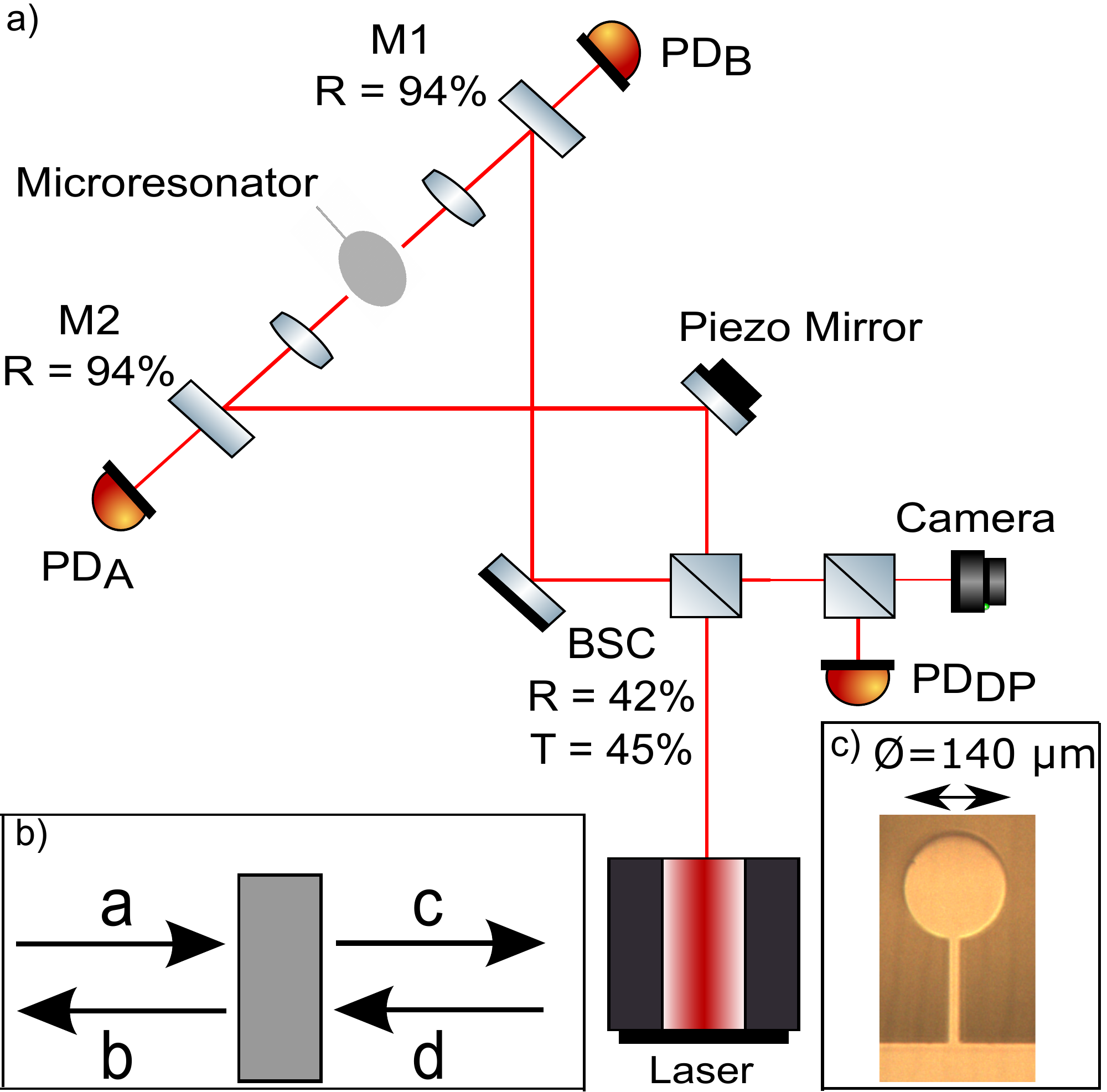}
\caption{Schematic of the experimental optomechanical setup. The Nd:YAG laser is split into two beams by a beamsplitter cube (BSC) and directed towards the partially reflective microresonator. One arm of the interferometer contains a steering mirror attached to a piezoeletric device that generates a phase difference $\phi$ between the two arms of the interferometer. Both arms contain partially transmissive steering mirrors (M1 and M2) that allow some of the reflected and transmitted light to be detected for locking the interferometer. The two reflected and two transmitted beams from the microresonator interfere at the BSC and are detected by a photodetector (PD$_\mathrm{DP}$).
(b) Fields $a$ and $d$ are incident on the microresonator from opposite sides. The input fields are supplied by a laser of power $P_0$ and frequency $\omega_0$ with $a$ and $d$ each receiving about half the total power. The microresonator has power refelctivity $R_{\mathrm{osc}} = \rho^2 = 65\%$. (c) Photomicrograph of the microresonator with a diameter of 140 $\mu$m supported by a 200 $\mu$m long by 20  $\mu$m wide cantilever structure.}
\label{Schematic}
\end{figure}

The optomechanical setup is shown in Fig. \ref{Schematic}. The in-air Michelson-Sagnac interferometer contains a partially reflective microresonator as the end/common mirror of the interferometer. The Michelson-Sagnac topology was used to simplifiy the alignment of the laser beams onto the microresonator. The microresonator is similar to the one used in \cite{Singh_PRL, Cripe_RPL} and described in \cite{cole08, cole12} but consists only of a 358.1-nm thick GaAs cantilever without the highly reflective stack of crystalline Al$_{ 0.92}$Ga$_{0.08}$As/GaAs layers. The GaAs microresonator has a power reflectivity of $R_\mathrm{osc}=65 \%$ for the laser wavelength of $\lambda = 1064 \mathrm{nm}$. The microresonator has a diameter of 140 \textmu m, a mass of about 30 ng, a natural mechanical frequency of $\Omega_\mathrm{m}=2\pi\times 850$ Hz, and a quality factor of  the fundamental resonance
\begin{eqnarray} \label{Q}
Q_\mathrm{f} = \frac{\Omega_\mathrm{m}}{\Gamma_\mathrm{m_\mathrm{f}}  (\Omega)} = \frac{\Omega_\mathrm{m}}{1100 \times (\frac{\Omega}{\Omega_\mathrm{m}})^{0.3}},
\end{eqnarray}
which is obtained by matching the theoretical model to the measured data.
The large mechanical damping, $\Gamma_\mathrm{m_\mathrm{f}}$, is a result of performing the experiment in air. A photomicrograph of the microresonator is included as a subset in Fig. \ref{Schematic}c.

\section{Theory}
To realize the optical spring, let us first consider the microresonator and its associated normalized fields as shown in Fig. \ref{Schematic}b \cite{Corbitt_math}. The normalized  input fields $a$ and $d$ each receive half the power from the laser source, and we allow for a phase shift in $d$, accounting for the difference between the path lengths. We assume the motion of the microresonator is small and that the path length difference remains constant, so we relate the normalized fields:

\begin{eqnarray}
&a = \sqrt{\frac{P_0}{2}}\\ &b = \rho a + \tau d \\ &c = \tau a  -\rho d \\  &d =\sqrt{\frac{P_0}{2}} e^{i\phi} \\ &\phi = \frac{L \omega_0}{c},
\end{eqnarray}
where $\rho$ and $\tau$ are the amplitude reflectivity and trasmissitivity such that $\rho^2 + \tau^2 = 1$, $\omega_0$ is the laser frequency, $P_0$ is the laser power incident on the beamsplitter cube, and $L$ and $\phi$ are the difference in length and phase of the two interferometer arms, respectively. We solve the equations and find the net power leaving the microresonator 

\begin{equation} \label{Power}
P_\mathrm{net} = |b|^2 - |c|^2 = 2 \rho \tau P_0 \cos\phi.
\end{equation}

To understand why we are interested in the net power leaving the microresonator, consider the forces acting on the microresonator. The net force from $a$, $b$, $c$, and $d$ is
\begin{equation} \label{F_net}
F_\mathrm{net} = (P_a + P_b -  P_c -  P_d) / c.
\end{equation}
If the input powers $P_a$ and $P_d$ are balanced, then a nonzero value for $P_b -  P_c$ gives rise to a net force on the microresonator exerted by radiation pressure $F_\mathrm{RP}= P_\mathrm{net}/c$. For small displacements $\delta L$ around an equilibrium position, the microresonator experiences a differential force

\begin{equation} \label{F_RP}
\delta F_\mathrm{RP} = \frac{1}{c} \frac{dP_\mathrm{net}}{dL}\delta L
\end{equation}
which can be expressed as an equivalent spring constant

\begin{eqnarray} \label{K_OS}
 K_\mathrm{OS} &=& -\frac{1}{c} \frac{dP_\mathrm{net}}{dL} = \frac{2}{c^2} \omega P_0 \rho \tau \sin(\phi).
\end{eqnarray}
 K$_\mathrm{OS}$ is purely real indicating that it provides a restoring force without the addition of a damping force. The maximum K$_\mathrm{OS}$ occurs for a path difference of $\phi = \sfrac{\pi}{2}$, as shown in Fig. \ref{Power_phi}.

\section{Experiment}

One of the arms of the interferometer contains a steering mirror which is mounted onto a piezoelectric device. The piezo mirror is used to control the phase difference between the two arms of the interferometer and to lock the interferometer. The steering mirrors on either side of the microresonator have a power reflectivity of 94\% to allow for some of the light to be used for locking the interferometer. The interferometer is locked by taking the signal from either $\mathrm{PD_{DP}}$, $\mathrm{PD}_A$, or $\mathrm{PD}_B$, filtering it, and feeding it back to the piezo mounted to the mirror in one of the arms of the interferometer. The relative phase difference between the two interferometer arms can be adjusted by tuning the locking setpoint on the PID controller.

\section{Data and Discussion}

We measure the optical spring effect by measuring the optical response of the system. This is performed by modulating the piezo in the interferometer, and measuring the resulting power fluctuation at one of the photodetectors as a function of the modulating frequency. In the absence of an optical spring, we should measure a featureless response. However, with an optical spring, we measure the effective closed loop gain of the optomechanical system
\begin{eqnarray} \label{G_CL}
G_{\mathrm{cl}} &=& \frac{1}{1+G_\mathrm{OS}} \\
&=& \frac{\Omega_\mathrm{m_f}^2-\Omega^2+i\Omega\Gamma_\mathrm{m_f}(\Omega)}{\Omega_\mathrm{m_f}^2-\Omega^2+i\Omega\Gamma_\mathrm{m_f}(\Omega) + \Omega_{\mathrm{OS}}^2} \nonumber \\ 
&+&\frac{\Omega_\mathrm{m_y}^2-\Omega^2+i\Omega\Gamma_\mathrm{m_y}}{\Omega_\mathrm{m_y}^2-\Omega^2+i\Omega\Gamma_\mathrm{m_y}+ \Omega_{\mathrm{OS}}^2}
\end{eqnarray}
as described in \cite{Cripe_RPL} where the first term is for the fundamental mode and the second term is for the yaw mode with $\Gamma_\mathrm{m_y} = 2000$ Hz. 

\begin{figure}
\center
\includegraphics[width=0.5\columnwidth]{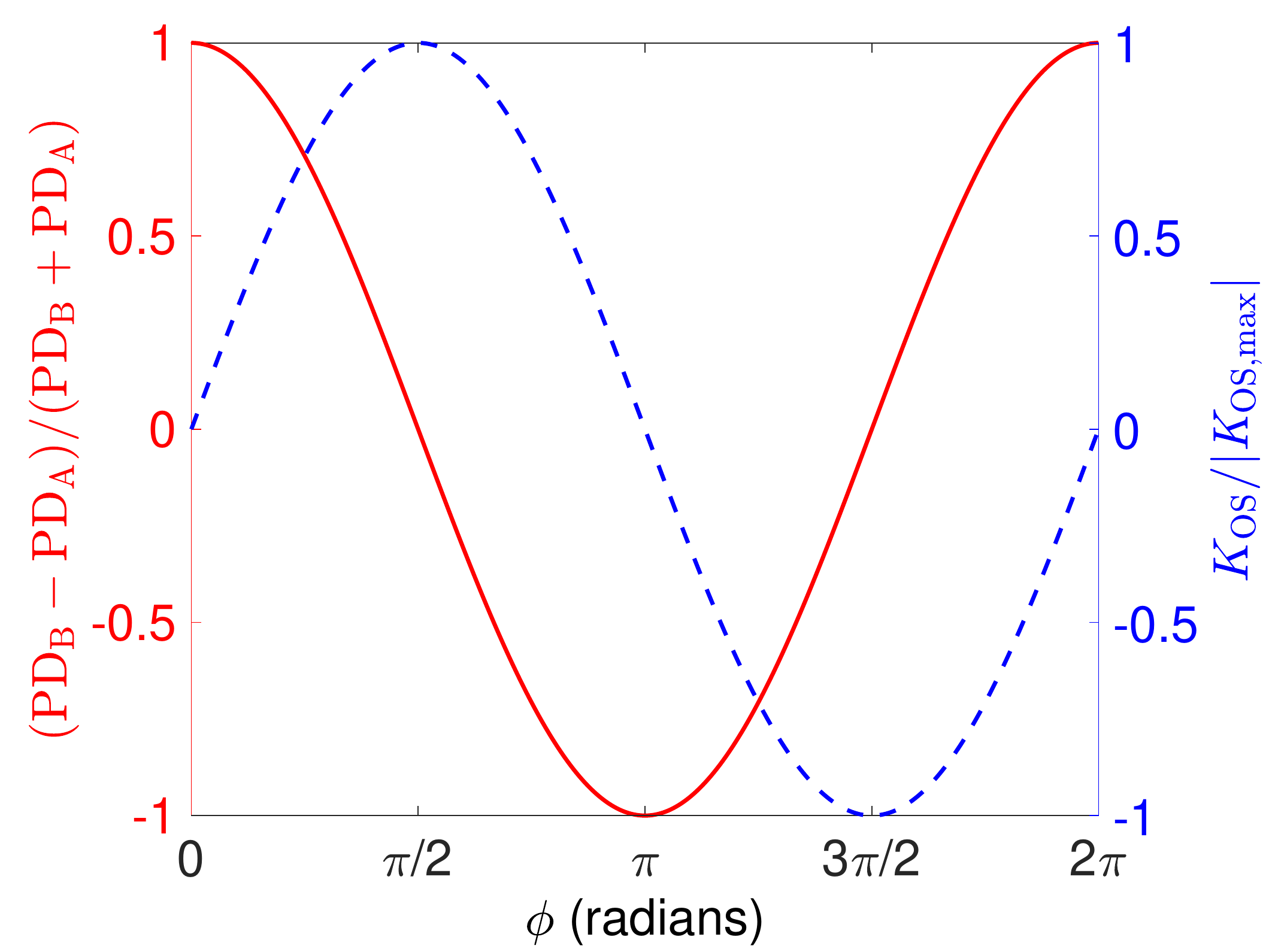}
\caption{Plot of the normalized power at the side photodetectors  and of the normalized K$_\mathrm{OS}$ as a function of $\phi$. The interferometer is locked at approximately $\phi = \sfrac{\pi}{2}$ where the optical spring effect is largest.}
\label{Power_phi}
\end{figure}

\begin{figure}[h!]
\center
\includegraphics[width=.5\columnwidth]{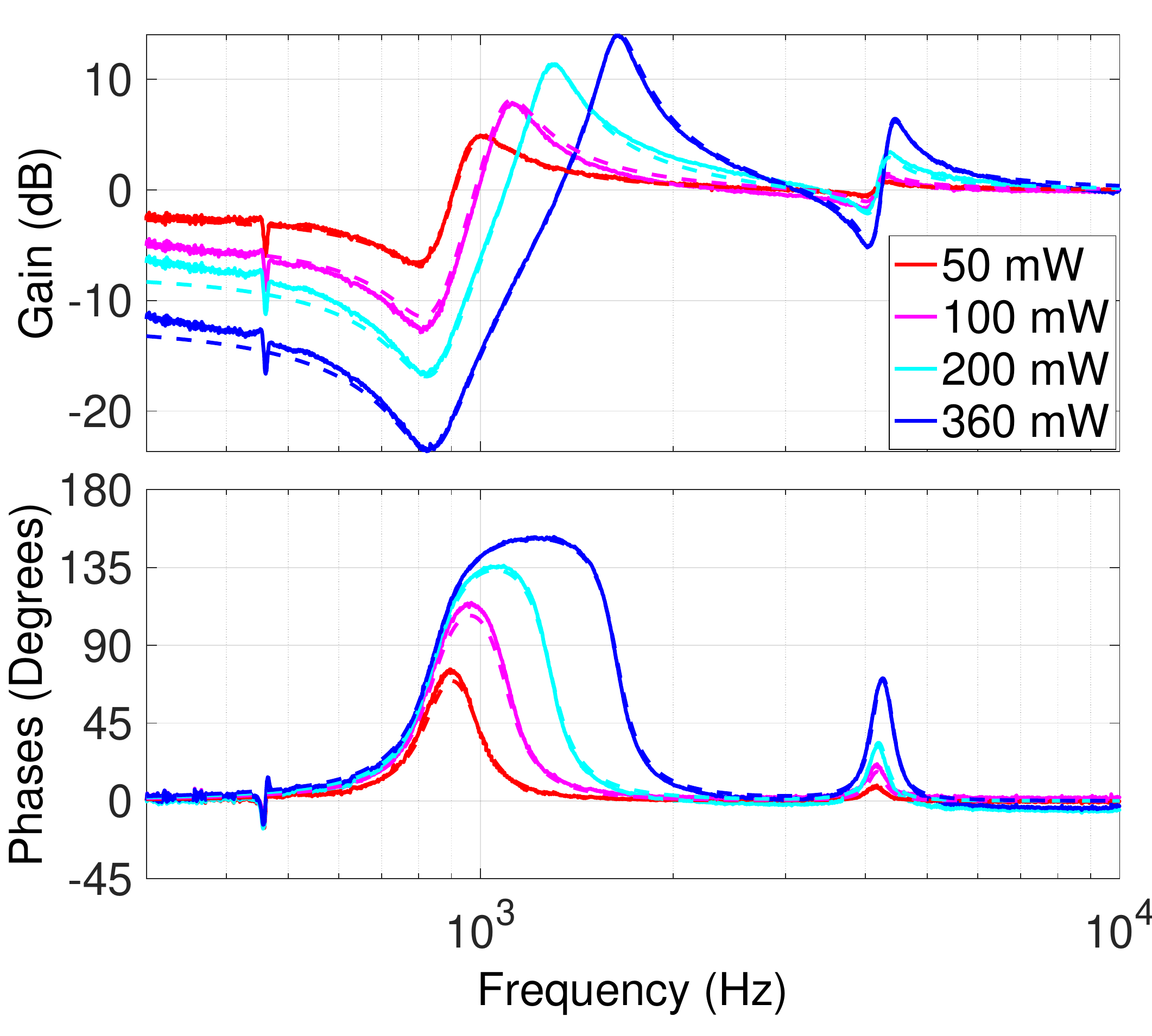}
\caption{Measurements (solid) and theoretical model (dashed) of the optical spring at input powers of 50 mW, 100 mW, 200 mW, and 360 mW. The measured transfer function is taken by injecting a signal to the PID controller connected to the piezo mirror in one arm of the interferometer and measuring its effect at $\mathrm{PD_{DP}}$. The dip at about 850 Hz corresponds to the fundamental mechanical resonance of the microresonator, and the feature at 4.2 kHz is the optical spring coupled to the yaw mode of the microresonator.}
\label{OS_measurements}
\end{figure}

We lock the interferometer at the mid-fringe point of $\mathrm{PD}_B$, which corresponds to the point at which the optical spring is largest, as shown in Eqs. \ref{Power} and \ref{K_OS}, and in Fig. \ref{Power_phi}. We measure the transfer function at input powers of 50 mW, 100 mW, 200 mW, and 360 mW, as shown in Fig. \ref{OS_measurements}. The optical spring peak is visible in each of the measurements at frequencies of 1000 Hz, 1120 Hz, 1310 Hz, and 1640 Hz, as well as a dip corresponding to the fundamental mechanical resonance at about 850 Hz. The effect of the optical spring is also visible on the yaw mode of the microresonator at at 4.2 kHz.

An interesting feature of the system is its ability to remain locked without any external feedback. At all four input powers, the optical spring is strong enough to stabilize the system and keep the interferometer locked at a desired fringe setpoint without the application of any feedback. Unlike the traditional case of the optical spring in a detuned Fabry-P\'erot cavity where the antidamping of the optical spring must be controlled using electronic feedback or another method, our system does not have an antidamping term and is therefore stable as a result of the restoring force provided by the optical spring.
External disturbances at frequencies below the optical spring frequency are suppressed by a factor of approximately 
\begin{eqnarray}
\frac{1}{G_\mathrm{cl}} \approx \frac{\Omega_\mathrm{OS}^2}{\Omega_\mathrm{m}^2} \approx 4
\end{eqnarray}
for the 360 mW measurement at low frequencies \cite{9, Cripe_RPL}. 
The stability of the system is visible in Fig. \ref{OS_measurements} where the noise at 300 Hz is suppressed by a factor of up to 11.6 dB or a magnitude of approximately 4 \footnote{The measurements shown in Fig. 3 are taken with the interferometer locked with the PID controller to avoid overly exciting a resonance and causing the system to lose lock}.  Further suppression of the external disturbances could be achieved by increasing the optical spring frequency by increasing the input power.

\section{Conclusion}
In conclusion, we have shown the measurement of the optical spring from a beamsplitter in a Michelson-Sagnac interferometer, without the use of a cavity. The measurements at input powers of 50 mW, 100 mW, 200 mW, and 360 mW clearly show the change in the system's resonance frequency created by the optical spring effect and match well with theoretical predictions. The optical spring created at all four input powers is strong enough to keep the interferometer stable and locked to the desired fringe setpoint and reduces disturbances at 300 Hz by up to 11.6 dB. 

In the future, we would like to investigate the possibility of using the partially reflective microresonators in experiments with quantum radiation pressure noise. As a result of not having the highly reflective stack of crystalline Al$_{ 0.92}$Ga$_{0.08}$As/GaAs layers, the mass of these microresonators is lower than the highly reflective microresonators. The reduction in mass, $m$, increases the signal-to-noise ratio of the quantum radiation pressure noise over the thermal noise by a factor of $\sqrt{m}$. We also aim to measure the mechanical dissipation as a function of frequency to investigate thermal noise models. In addition, the microresonators could have use in experiments studying unstable optomechanical filter cavities, such as those proposed in \cite{Haixing_FC}.




This work was funded by National Science Foundation CAREER grant PHY-1150531.

This document has been assigned the LIGO document number LIGO-P1700377. 




\begin{thebibliography}{}
\newcommand{\enquote}[1]{``#1''}

\end{thebibliography}


\begin{thebibliography}{10}
\newcommand{\enquote}[1]{``#1''}

\bibitem{24}
M.~Aspelmeyer, T.~J. Kippenberg, and F.~Marquardt, Rev. Mod. Phys. \textbf{86},
  1391 (2014).

\bibitem{Braginsky_1}
V.~B. Braginsky and I.~I. Minakova, Moscow Univ. Phys. Bull. \textbf{1}, 83
  (1964).

\bibitem{Braginsky_2}
V.~B. Braginsky and A.~B. Manukin, Soviet Physics JETP \textbf{25}, 653 (1967).

\bibitem{13}
B.~S. Sheard, M.~B. Gray, C.~M. Mow-Lowry, D.~E. McClelland, and S.~E.
  Whitcomb, Phys. Rev. A \textbf{69}, 051801 (2004).

\bibitem{9}
T.~Corbitt, Y.~Chen, F.~Khalili, D.~Ottaway, S.~Vyatchanin, S.~Whitcomb, and
  N.~Mavalvala, Phys. Rev. A \textbf{73}, 023801 (2006).

\bibitem{Cripe_RPL}
J.~Cripe, N.~Aggarwal, R.~Singh, R.~Lanza, A.~Libson, M.~J. Yap, G.~D. Cole,
  D.~McClelland, N.~Mavalvala, and T.~Corbitt, In Preparation .

\bibitem{2-2}
O.~Miyakawa, R.~Ward, R.~Adhikari, M.~Evans, B.~Abbott, R.~Bork, D.~Busby,
  J.~Heefner, A.~Ivanov, M.~Smith, R.~Taylor, S.~Vass, A.~Weinstein,
  M.~Varvella, S.~Kawamura, F.~Kawazoe, S.~Sakata, and C.~Mow-Lowry, Phys. Rev.
  D \textbf{74}, 022001 (2006).

\bibitem{15}
T.~Corbitt, D.~Ottaway, E.~Innerhofer, J.~Pelc, and N.~Mavalvala, Phys. Rev. A
  \textbf{74}, 021802 (2006).

\bibitem{16}
T.~J. Kippenberg, H.~Rokhsari, T.~Carmon, A.~Scherer, and K.~J. Vahala, Phys.
  Rev. Lett. \textbf{95}, 033901 (2005).

\bibitem{17}
T.~Corbitt, Y.~Chen, E.~Innerhofer, H.~M\"uller-Ebhardt, D.~Ottaway,
  H.~Rehbein, D.~Sigg, S.~Whitcomb, C.~Wipf, and N.~Mavalvala, Phys. Rev. Lett.
  \textbf{98}, 150802 (2007).

\bibitem{Singh_PRL}
R.~Singh, G.~D. Cole, J.~Cripe, and T.~Corbitt, Phys. Rev. Lett. \textbf{117},
  213604 (2016).

\bibitem{18}
D.~Kelley, J.~Lough, F.~Manga\~na Sandoval, A.~Perreca, and S.~W. Ballmer,
  Phys. Rev. D \textbf{92}, 062003 (2015).

\bibitem{PA_thermal}
P.~A. Altin, T.~T.-H. Nguyen, B.~J.~J. Slagmolen, R.~L. Ward, D.~A. Shaddock,
  and D.~E. McClelland, Scientific Reports \textbf{7}, 14546 (2017).

\bibitem{LIGO}
J.~A. et~al., Classical and Quantum Gravity \textbf{32}, 074001 (2015).

\bibitem{VIRGO}
F.~A. et~al., Classical and Quantum Gravity \textbf{32}, 024001 (2015).

\bibitem{Chen}
A.~Buonanno and Y.~Chen, Classical and Quantum Gravity \textbf{19}, 1569
  (2002).

\bibitem{1-1}
A.~Buonanno and Y.~Chen, Phys. Rev. D \textbf{65}, 042001 (2002).

\bibitem{25}
A.~Xuereb, R.~Schnabel, and K.~Hammerer, Phys. Rev. Lett. \textbf{107}, 213604
  (2011).

\bibitem{19}
A.~Sawadsky, H.~Kaufer, R.~M. Nia, S.~P. Tarabrin, F.~Y. Khalili, K.~Hammerer,
  and R.~Schnabel, Phys. Rev. Lett. \textbf{114}, 043601 (2015).

\bibitem{Danilishin2012}
S.~L. Danilishin and F.~Y. Khalili, Living Reviews in Relativity \textbf{15}, 5
  (2012).

\bibitem{cole08}
G.~D. Cole, S.~Gr{\"o}blacher, K.~Gugler, S.~Gigan, and M.~Aspelmeyer, Applied
  Physics Letters \textbf{92}, 261108 (2008).

\bibitem{cole12}
G.~D. Cole, \enquote{Cavity optomechanics with low-noise crystalline mirrors,}
  in \enquote{Proc. SPIE 8458, Optics \& Photonics, Optical Trapping and
  Optical Micromanipulation IX,}  (SPIE, 2012), p. 845807.

\bibitem{Corbitt_math}
T.~Corbitt, Y.~Chen, and N.~Mavalvala, Phys. Rev. A \textbf{72}, 013818 (2005).

\bibitem{Haixing_FC}
H.~Miao, Y.~Ma, C.~Zhao, and Y.~Chen, Phys. Rev. Lett. \textbf{115}, 211104
  (2015).

\end{thebibliography}

\end{document}